\documentstyle[12pt,epsfig]{article}
\begin{document}
\noindent
\begin{center}
{\Large {\bf Notes on the Chameleon Brans-Dicke Gravity}}\\
\vspace{2cm}
 ${\bf Yousef~Bisabr}$\footnote{e-mail:~y-bisabr@srttu.edu.}\\
\vspace{.5cm} {\small{Department of Physics, Shahid Rajaee Teacher
Training University,
Lavizan, Tehran 16788, Iran}}\\
\end{center}
\vspace{1cm}
\begin{abstract}
We consider a generalized Brans-Dicke model in which the scalar field has a potential function and is also allowed to couple
non-minimally with the matter sector.  This anomalous gravitational coupling can in principle avoid the model to pass local gravity experiments.  One then usually assumes that
the scalar field has a chameleon behavior in the sense that it acquires a density-dependent effective mass.  While it can take a small effective mass in cosmological (low-density environment)
scale, it has a sufficiently heavy mass in Solar System (large-density environment) and then hides gravity tests.  We will argue that such a chameleon behavior can
not be generally realized and depends significantly on the forms attributed to the potential and the coupling functions.

\end{abstract}
~~~~~~~PACS Numbers: 04.50.Kd, 04.20.Cv, 95.36.+x \vspace{3cm}
\section{Introduction}
One of the approaches to explain accelerating expansion of the
universe is to attribute this phenomenon to some modifications of
general relativity. Such modified gravity models can be obtained in
different ways. For instance, one can replace the Ricci scalar in
the Einstein-Hilbert action by some functions $f(R)$ (for a review
see, e.g., \cite{1} and references therein), or by considering a
scalar partner for the metric tensor for describing geometry of
spacetime, the so-called scalar-tensor gravity.  The prototype of
the latter is Brans-Dicke (BD) theory \cite{BD} which its original
motivation was the search for a theory containing Mach's principle.
As the simplest and best-studied generalization of general
relativity, it is natural to think about the BD scalar field as a
possible candidate for producing cosmic acceleration without
invoking auxiliary fields or exotic matter systems. In fact, there
have been many attempts to show that BD model can potentially
explain the cosmic acceleration. It is shown that this theory can
actually produce a non-decelerating expansion for low negative
values of the BD parameter \cite{ban}. Unfortunately, this conflicts
with the lower bound imposed on this parameter by solar system
experiments \cite{will}.  Due to this difficulty, some authors
propose modifications of the BD model such as introducing some
potential functions for the scalar
field \cite{ban1}, or considering a field-dependent BD parameter \cite{ban2} without resolving the problem.\\
In a general scalar-tensor theory there is a non-minimal coupling
between the scalar field and Ricci scalar while the former minimally
couples with the matter sector.  In other terms, there is no an
explicit coupling between the scalar field and matter systems in
Jordan frame representation.  In BD theory, in its original form,
the motivation for such a minimal coupling was to keep the theory in
accord with the weak equivalence principle \cite{BD}. There has
recently been a tendency in the literature to go a step further and
consider a non-minimal coupling between the scalar field and matter
systems as well by introducing an arbitrary function of the scalar
field as a coupling function \cite{ban3} \cite{fan} \cite{bis}. In
these models, the scalar field is regarded as a chameleon field and
it is assumed that it can be heavy enough in the environment of the
laboratory tests so that the local gravity constraints suppressed.
Meanwhile, it can be light enough in the low-density cosmological
environment to be considered as a candidate for dark energy.  The
large scale behavior of this chameleon BD theory has been already
studied in a different work \cite{bis}.  It is clear that these
different behaviors in small and large scales depend crucially on
the shapes of the potential and the coupling functions since both
functions contribute to the effective mass or compton wavelength of
the scalar field. In the present work we will study behavior of the
theory in small scales.  We will study the conditions that should
hold in order that the theory passes solar system experiment. We
will show that the conditions can not be satisfied for some usual forms of
the potential and the coupling functions.
~~~~~~~~~~~~~~~~~~~~~~~~~~~~~~~~~~~~~~~~~~~~~~~~~~~~~~~~~~~~~~~~~~~~~~~~~~~~~~~~~~~~~~~~~~~~~~~~~~~~~
\section{The model}
We consider the action functional
\begin{equation}
S=\frac{1}{16\pi}\int d^4x \sqrt{-\bar{g}} \{\phi
\bar{R}-\frac{\omega}{\phi}\bar{g}^{\mu\nu}\bar{\nabla}_{\mu}\phi
\bar{\nabla}_{\nu}\phi -V(\phi)+16\pi f(\phi)L_m\}
\label{1}\end{equation} where $R$ is the Ricci scalar, $\phi$ is the
BD scalar field, $V(\phi)$ and $f(\phi)$ are some analytic
functions.  Here the matter Lagrangian density, denoted by $L_m$, is
coupled with $\phi$ via the function $f(\phi)$. This allows a
non-minimal interaction between the matter system and $\phi$. Taking
$f(\phi)=1$, we return to the BD action with
a potential function $V(\phi)$.  \\
A conformal transformation
\begin{equation}
\bar{g}_{\mu\nu}\rightarrow g_{\mu\nu}=\Omega^2 \bar{g}_{\mu\nu}
\label{a2}\end{equation} with $\Omega=\sqrt{G \phi}$ brings the
above action into the Einstein frame \cite{far1} \cite{far}.  Then a
scalar field redefinition
\begin{equation}
\varphi(\phi)=\sqrt{\frac{2\omega+3}{16\pi G}}\ln
(\frac{\phi}{\phi_0}) \label{a3}\end{equation} with $\phi_0\sim
G^{-1}$, $\phi>0$ and $\omega>-\frac{3}{2}$ transforms the kinetic
term of the scalar field into a canonical form.  In terms of the
variables ($g_{\mu\nu}$, $\varphi$) the action (\ref{1}) takes the
form
\begin{equation}
S_{EF}= \int d^{4}x \sqrt{-g} \{\frac{R}{16\pi G}
-\frac{1}{2}g^{\mu\nu}\nabla_{\mu}\varphi
\nabla_{\nu}\varphi-U(\varphi)+ \exp(-8\sqrt{\frac{\pi
G}{2\omega+3}}\varphi)~f(\varphi) L_{m}\} \label{a5}\end{equation}
Here $\nabla_{\mu}$ is the covariant derivative of the rescaled
metric $g_{\mu\nu}$.  The Einstein frame potential is given by
\begin{equation}
U(\varphi)= V(\phi(\varphi))~\exp(-\sigma\varphi/M_{p})
\label{a6}\end{equation} in which $\sigma=8\sqrt{\frac{\pi
}{2\omega+3}}$ and $M_p= G^{-1/2}$.\\
Varying the action (\ref{a5}) with respect to the metric
$g_{\mu\nu}$ and $\varphi$ yields the field equations,
\begin{equation}
G_{\mu\nu}=8\pi G (h(\varphi)T^{m}_{\mu\nu}+T^{\varphi}_{\mu\nu})
\label{2}\end{equation}
\begin{equation}
\Box \varphi-U'(\varphi)=-h'(\varphi) L_m \label{3}\end{equation}
where
\begin{equation}
T^{\varphi}_{\mu\nu}=(\nabla_{\mu}\varphi
\nabla_{\nu}\varphi-\frac{1}{2} g_{\mu\nu}\nabla_{\alpha}\varphi
\nabla^{\alpha}\varphi)
-U(\varphi)g_{\mu\nu} \label{4}\end{equation}
\begin{equation}
T^m_{\mu\nu}=\frac{-2}{\sqrt{-g}}\frac{\delta (\sqrt{-g}L_m)}{\delta
g^{\mu\nu}} \label{5}\end{equation} Here
$h(\varphi)=e^{-\sigma\varphi/M_{p}}f(\varphi)$,
$T^m=g^{\mu\nu}T^m_{\mu\nu}$ and prime indicates differentiation
with respect to $\varphi$. Due to explicit coupling of the matter
system with the scalar field, the stress-tensor $T^m_{\mu\nu}$ is
not divergence free. This can be seen by applying the Bianchi
identities $\nabla^{\mu}G_{\mu\nu}=0$ to (\ref{2}), which leads to
\begin{equation}
\nabla^{\mu}T^m_{\mu\nu}=(L_m-T^m)\nabla_{\nu}\ln h(\varphi)
\label{6}\end{equation} As it is clear from (\ref{6}), details of
the energy exchange between matter and $\varphi$ depends on the
explicit form of the matter Lagrangian density $L_m$.  Here we
consider a perfect fluid energy-momentum tensor as a matter system
\begin{equation}
T^m_{\mu\nu}=(\rho_m+p_m)u_{\mu}u_{\nu}+p_mg_{\mu\nu}
\label{b1}\end{equation}
where $\rho_m$ and $p_m$ are energy density and pressure, respectively. The four-velocity of the fluid is denoted by $u_{\mu}$.\\
There are different choices for the perfect fluid Lagrangian density
which all of them leads to the same energy-momentum tensor and field
equations in the context of general relativity \cite{2} \cite{3}.
The two Lagrangian densities that have been widely used in the
literature are $L_m=p_m$ and $L_m=-\rho_m$ \cite {3a} \cite{4}
\cite{5}.  For a perfect fluid that does not couple explicitly to
the curvature (i.e., for $f(\varphi) = 1$), the two Lagrangian
densities $L_m =p_m$ and $L_m=-\rho_m$ are perfectly equivalent, as
discussed in \cite{4} \cite{5}. However, in the model presented here
the expression of $L_m$ enters explicitly the field equations and
all results strongly depend on the choice of $L_m$.  In fact, it is
shown that there is a strong debate about equivalency of different
expressions attributed to the Lagrangian density of a coupled
perfect fluid \cite{6}. Here we take $L_m=-\rho_m$ for the lagrangian density.  \\
We consider $T^m_{\mu\nu}$ as the stress-tensor of dust. In a static
and spherically symmetric spacetime the equation (\ref{3}) gives
\begin{equation}
\frac{d^2 \varphi}{dr^2}+\frac{2}{r}
\frac{d\varphi}{dr}=\frac{dV_{eff}(\varphi)}{d\varphi}
\label{d}\end{equation} where $r$ is distance from center of the
symmetry in the Einstein frame and
\begin{equation}
V_{eff}(\varphi)=\{V(\varphi)+\rho_m f(\varphi)\}e^{-\sigma
\varphi/M_p} \label{d1}\end{equation} To proceed further, we should
have the explicit form of the functions $V(\varphi)$ and
$f(\varphi)$.  In the present work, we take $f(\varphi)$ as an
exponential function $f(\varphi)= e^{l_2\varphi/M_p}$ with $l_2$
being a constant dimensionless parameter.  For the potential
function, we will consider the following cases :\\
\subsection{Exponential potentials }
We first consider the potential $V(\varphi)= e^{l_1\varphi/M_p}$.  In this case, the
effective potential takes the form
\begin{equation}
V_{eff}(\varphi)=e^{(l_1-\sigma)\varphi/M_p}+\rho_m
e^{\beta\varphi/M_p} \label{d2}\end{equation} where
$\beta=l_2-\sigma$.  One can find solutions for
$V'_{eff}(\varphi)=0$, which gives
\begin{equation}
\varphi_{min}=\frac{M_p}{l_1-l_2}\ln[{\frac{\beta\rho_m}{(\sigma-l_1)}}]
\label{dd}\end{equation} This is a local minimum if the following
condition is satisfied
\begin{equation}
V''_{eff}(\varphi_{min})=\frac{\rho_m}{M_p^2}~e^{\beta\varphi/M_p}~\beta(l_2-l_1)>0
\label{d50}\end{equation} For a spherically symmetric body with a
radius $r_{c}$ and constant energy densities $\rho_{in}$ ($r<
r_{c}$) and $\rho_{out}$ ($r> r_{c}$), there is a thin-shell
condition \cite{k}
\begin{equation}
\frac{\Delta
r_{c}}{r_{c}}=\frac{\varphi_{min(out)}-\varphi_{min(in)}}{6\beta
M_p\Phi_{c}}\ll 1 \label{l1}\end{equation} where $\Phi_{c}=M_{c}/8\pi
M_p^2 r_{c}$ is the Newtonian potential at $r=r_{c}$ with $M_{c}$ being
the mass of the body.  In this expression, $\varphi_{min(in)}$ and
$\varphi_{min(out)}$ denote the field values at two minima of the
effective potential $V_{eff}(\varphi)$ inside and outside the
object, respectively. They must clearly satisfy
$V^{'}_{eff}(\varphi_{min(in)})=0$ and
$V^{'}_{eff}(\varphi_{min(out)})=0$. In this case, equation
(\ref{d}) with some appropriate boundary conditions gives the field
profile outside the object \cite{k}
\begin{equation}
\varphi(r)=-\frac{\beta}{4\pi M_p} \frac{3\Delta
r_{c}}{r_{c}}\frac{M_{c} e^{-m_{out}
(r-r_{c})}}{r}+\varphi_{min(out)} \label{l2}\end{equation} As usual,
masses of small fluctuations about the minima are given by
$m_{in}=[V^{''}_{eff}(\varphi_{min(in)})]^{\frac{1}{2}}$ and
$m_{out}=[V^{''}_{eff}(\varphi_{min(out)})]^{\frac{1}{2}}$ which
depend on ambient matter density. A region with large mass density
corresponds to a heavy mass field while regions with low mass
density corresponds to a field with lighter mass. In this way it is
possible for the mass field to take sufficiently large values near
massive objects in the Solar System scale and to hide the local
tests.\\
$1.~ Thin-shell~~ condition$\\
In the chameleon mechanism, the chameleon field is trapped inside
large and massive bodies and its influence on the other bodies is
only due to a thin-shell near the surface of the body.  The
criterion for this thin-shell condition is given by (\ref{l1}). If
we combine (\ref{dd}) and (\ref{l1}) we obtain
\begin{equation}
\frac{\Delta r_{c}}{r_{c}}=\frac{1}{6\Phi_c \beta(l_1-l_2)
}\ln{\frac{\rho_{out}}{\rho_{in}}} \label{hh}\end{equation}
 In
weak field approximation, the spherically symmetric metric in the
Jordan frame is given by
\begin{equation}
ds^{2}=-[1-2X(\bar{r})]dt^{2}+[1+2Y(\bar{r})]d\bar{r}^{2}+\bar{r}^2
d\Omega^2
\end{equation}
where $X(\bar{r})$ and $Y(\bar{r})$ are some functions of $\bar{r}$.
There is a relation between $r$ and $\bar{r}$ so that $r=\Omega
\bar{r}$ with $\Omega=e^{\sigma\varphi/4M_p}$.  Note that local
gravity experiments constrain the BD parameter so that $\omega>3500$
\cite{will}, or equivalently $\sigma<0.17$.  For $\varphi/M_p$ not
much greater than unity, it implies that $\bar{r} \approx r$.
Assuming $ m_{out}~r\ll 1 $, namely that the Compton wavelength
$m_{out}^{-1}$ is much larger than Solar System scales, the
chameleon mechanism gives for the post-Newtonian parameter $\gamma$
\cite{faulk}
\begin{equation}
\gamma=\frac{3-\frac{\Delta r_{c}}{r_{c}}}{3+\frac{\Delta
r_{c}}{r_{c}}} \simeq 1-\frac{2}{3}\frac{\Delta r_{c}}{r_{c}}
\label{gg}\end{equation}\\
We can now apply (\ref{hh}) on the Earth and obtain the condition
that the Earth has a thin-shell. To do this, we assume that the
Earth is a solid sphere of radius $R_{e} =6.4\times 10^{8}~cm$ and
mean density $\rho_{e} \sim 10~gr/cm^{3}$. We also assume that the
Earth is surrounded by an atmosphere with homogenous density
$\rho_{a} \sim 10^{-3}~gr/cm^{3}$ and thickness $100 km$.  In this
case, (\ref{hh}) takes the form
\begin{equation}
\frac{\Delta R_{e}}{R_{e}}=\frac{1}{6\Phi_e \beta(l_1-l_2)
}\ln{\frac{\rho_{a}}{\rho_{e}}} \label{hh1}\end{equation} in which
 $\Phi_{e}=6.95\times
10^{-10}$ is Newtonian potential on surface of the Earth
\cite{wein}.  The tightest Solar System constraint on $\gamma$ comes
from Cassini tracking which gives $\mid \gamma -1 \mid < 2.3 \times
10^{-5}$ \cite{will}. This together with (\ref{gg}) and (\ref{hh1})
yields
\begin{equation} |(l_1-l_2)(l_2-\sigma)|>   10^{14}\label{hh2}\end{equation}
2. $Equivalence ~ principle$\\
We now consider constraints coming from possible violation of weak
equivalence principle. We assume that the Earth, together with its
surrounding atmosphere, is an isolated body and neglect the effect
of the other compact objects such as the Sun, the Moon and the other
planets. Far away the Earth, matter density is modeled by a
homogeneous gas with energy density $\rho_{G}\sim 10^{-24}
gr/cm^{3}$.  To proceed further, we first consider the condition
that the atmosphere of the Earth satisfies the thin-shell condition
\cite{k}. If the atmosphere has a thin-shell the thickness of the
shell ($\Delta R_{a}$) must be clearly smaller than that of the
atmosphere itself, namely $\Delta R_{a}< R_{a}$, where $R_{a}$ is
the outer radius of the atmosphere. If we take thickness of the
shell equal to that of the atmosphere itself $\Delta R_{a}\sim
10^{2}~km$ we obtain $\frac{\Delta R_{a}}{R_{a}}<1.5\times 10^{-2}$.
It is then possible to relate $\frac{\Delta
R_{e}}{R_{e}}=\frac{\varphi_{min(a)}-\varphi_{min(e)}}{6\beta M_p
\Phi_{e}}$ and $\frac{\Delta
R_{a}}{R_{a}}=\frac{\varphi_{min(G)}-\varphi_{min{a}}}{6\beta M_p
\Phi_{a}}$ where $\varphi_{min(e)}$, $\varphi_{min(a)}$ and
$\varphi_{min(G)}$ are the field values at the local minimum of the
effective potential in the regions $r<R_{e}$ , $R_{a}>r>R_{e}$ and
$r>R_{a}$ respectively. Using the fact that newtonian potential
inside a spherically symmetric object with mass density $\rho$ is
$\Phi \propto\rho R^{2}$, one can write $\Phi_{e}=10^{4}~\Phi_{a}$
where $\Phi_{e}$ and $\Phi_{a}$ are Newtonian potentials on the
surface of the Earth and the atmosphere, respectively. This gives
$\Delta R_{e}/ R_{e} \approx 10^{-4}~\Delta R_{a}/ R_{a}$. With
these results, the condition for the atmosphere to have a thin-shell
is
\begin{equation}
\frac{\Delta R_{e}}{R_{e}} < 1.5\times 10^{-6}
\label{R}\end{equation}\\
The tests of equivalence principle measure the difference of
free-fall acceleration of the Moon and the Earth towards the Sun.
The constraint on the difference of the two acceleration is given by
\cite{will}
\begin{equation}
\frac{|a_{m}-a_{e}|}{a_{N}} < 10^{-13} \label{f}\end{equation} where
$a_{m}$ and $a_{e}$ are acceleration of the Moon and the Earth
respectively and $a_{N}$ is the Newtonian acceleration.  The Sun and
the Moon are all subject to the thin-shell condition \cite{k} and
the field profile outside the spheres are given by (\ref{l2}) with
replacement of corresponding quantities.  The accelerations $a_{m}$
and $a_{e}$ are then given by \cite{k}
\begin{equation}
a_{e}\approx a_{N}\{1+18\beta^2 (\frac{\Delta R_{e}}{R_{e}})^2
\frac{\Phi_{e}}{\Phi_{s}}\}
\end{equation}
\begin{equation}
a_{m}\approx a_{N}\{1+18\beta^2 (\frac{\Delta R_{e}}{R_{e}})^2
\frac{\Phi_{e}^2}{\Phi_{s}\Phi_{m}}\}
\end{equation}
where $\Phi_{e}=6.95\times 10^{-10}$, $\Phi_{m}=3.14\times 10^{-11}$
and $\Phi_{s}=2.12\times 10^{-6}$ are Newtonian potentials on the
surfaces of the Earth, the Moon and the Sun, respectively
\cite{wein}. This gives a difference of free-fall acceleration
\begin{equation}
\frac{|a_{m}-a_{e}|}{a_{N}}\approx \beta^2~(\frac{\Delta R_{e}}{R_{e}})^2
\end{equation}
Combining this with (\ref{f}) results in
\begin{equation}
\beta~\frac{\Delta R_{e}}{R_{e}} <  10^{-7} \label{RR}\end{equation}
  Taking this as the constraint coming
from violation of equivalence principle and combining with
(\ref{hh1}), we obtain
\begin{equation}
|l_1-l_2|>  10^{16} \label{nn1}\end{equation}\\\\
The constraints (\ref{hh2}) and (\ref{nn1}) imply that $l_2-\sigma
\sim 10^{-2}$ and one can take $l_2\sim \sigma<0.17$.  This means
that local experiments are satisfied only for extremely large values
of $l_1$.  Since $l_1>l_2$, the condition (\ref{d50}) and the expression (\ref{dd}) require that $\beta<0$.  Thus the second term in the effective potential
is a decreasing function while $U'(\varphi)>0$.  This ensures that $V_{eff}(\varphi)$ does exhibit a local minimum corresponding to an effective mass.

\subsection{Power-law potentials }In this case, we take $V(\varphi)=V_0\varphi^{l_3}$ with $l_3$ being a dimensionless
constant parameter.  The effective potential takes
then the form
\begin{equation}
%\frac{d^2 \varphi}{dr^2}+\frac{2}{r}
%\frac{d\varphi}{dr}=\frac{dU(\varphi)}{d\varphi}+\frac{\beta}{M_p}\rho_m
%e^{\beta\varphi/M_p}
V_{eff}(\varphi)=V_o\varphi^{l_3}e^{-\sigma\varphi/M_p}+\rho_m e^{\beta\varphi/M_p}
\label{d2}\end{equation}  For $\varphi<<M_p$,
one can write $e^{l_2\varphi/M_p}\approx
1+\frac{l_2\varphi}{M_p}$\footnote{ Here, we assume that $l_2$ is
not much greater than unity.}. Then one can find solutions for
$V'_{eff}(\varphi)=0$, which gives
\begin{equation}
 V_0
\varphi^{l_3}(\frac{l_3}{\varphi}-\frac{\sigma}{M_p})-\frac{\beta}{M_p}\rho_m(1+\frac{l_2\varphi}{M_p})=0
\label{d3}\end{equation} Since $\sigma<<1$, one can write
$\sigma\varphi<<<M_p$ which means that the second term in the first
parentheses can be neglected.  In this case, (\ref{d3}) gives for
$l_3=2$,
\begin{equation}
\varphi_{min}=\frac{\beta\rho_m/M_p}{2V_0-l_2\beta\rho_m/M_p^2}
\label{d4}\end{equation} This is a local minimum if the following
condition is satisfied
\begin{equation}
V''_{eff}(\varphi_{min})=2V_0-\frac{l_2\beta}{M_p^2}\rho_m>0
\label{d5}\end{equation}  If we combine (\ref{d4}) and (\ref{l1}) we
obtain
\begin{equation}
\frac{\Delta r_{c}}{r_{c}}=\frac{V_0}{3\Phi_c \beta^2
l_2^2}\frac{M_p^2}{\rho_{out}} \label{h}\end{equation} where
 we have used that facts that
$\rho_{out}<<\rho_{in}$ and $M_p^2<<\rho_{in}$ and $\rho_{out}$. For
Earth to have a thin-shell, (\ref{h}) takes the form
\begin{equation}
\frac{\Delta R_{e}}{R_{e}}=\frac{1}{3\Phi_e \beta^2 l_2^2
}\frac{M_p^2}{\rho_{a}} \label{h1}\end{equation} in which $V_0$ is
taken to be of order of unity. The bound on the PPN parameter
$\gamma$ then reads\footnote{We have used $M_p^2/\rho_a\sim
10^{-36}~cm^{-2}$.}
\begin{equation} |l_2(\sigma-l_2)|>   10^{-12}\label{h2}\end{equation}
%The constraint coming from equivalence principle can be obtained by
%combining (\ref{RR}) with (\ref{h1}) which gives
%\begin{equation}
%|l_2^2(\sigma-l_2)|>  10^{-20} \label{nn}\end{equation}
Let us compare the latter constraint with (\ref{d5}).  We first note
that $\rho_{out}/ M_p^2\sim 10^{36}~cm^{2}$ and $\rho_{in}/
M_p^2\sim 10^{40}~cm^{2}$.  For $V_0\sim 1$, (\ref{d5}) implies
that $l_2(\sigma-l_2)<10^{-36}$ or $10^{-40}$ which is not
consistent with (\ref{h2}).  It is also possible to translate (\ref{d5}) into a constraint on the scale introduced by $V_0$ for reasonable values of
$l_2$.  In this case, (\ref{d5}) and (\ref{h2}) still remain inconsistent.
~~~~~~~~~~~~~~~~~~~~~~~~~~~~~~~~~~~~~~~~~~~~~~~~~~~~~~~~~~~~~~~~~~~~~~~~~~~~~~~~~~~~~~~~~~~~~~~~~~~~~~~~~~~~~~~~~~~~~~~~~~~~~~~~~~~~~~~~~~~
\section{Conclusions}
In chameleon BD gravity, the BD scalar field is allowed to couple with matter sector via an arbitrary coupling function.
One then pre-assumes that the scalar field is a chameleon in the sense that it can hide the anomalous coupling
via chameleon mechanism and pass local gravity experiments. In this note, we have checked viability of this pre-assumption in
some extent. To do this, we have written the model in the Einstein conformal frame.  In Einstein frame
representation, the matter system couples with the scalar field via
two different functions, one exponential function which is given by the conformal transformation
and the other $f(\varphi)$ which is also
assumed to be an exponential function parameterized by $l_2$.  We
have considered conditions that the whole anomalous coupling is
suppressed by the chameleon mechanism.\\
When $V(\varphi)= e^{l_1 \varphi/M_p}$, the thin-shell condition
for the earth gives $l_2\sim \omega^{-1/2}$.  On the other hand, the equivalence principle sets
a lower bound $l_1>10^{16}$ which implies that the
local gravity experiments can not be suppressed for reasonable values
of the exponent.
We have also examine power law potentials $V(\varphi)=V_0\varphi^{l_3}$.  For a quadratic potential ($l_3=2$), we have obtained a certain condition
on the parameter $l_2$ for which the potential has a local minimum.  We have shown that this condition is not consistent with the thin-shell condition for the earth.
These results create strong debates over viability of the pre-assumed chameleon behavior of the scalar field in the context of chameleon BD gravity models.  This behavior seems
to depend significantly on the potential and the coupling functions.

~~~~~~~~~~~~~~~~~~~~~~~~~~~~~~~~~~~~~~~~~~~~~~~~~~~~~~~~~~~~~~~~~~~~~~~~~~~~~~~~~~~~~~~~~~~~~


\begin{thebibliography}{99}
\bibitem{1}T. P. Sotiriou and V. Faraoni, Rev. Mod. Phys. {\bf 82}, 451 (2010)
\bibitem{BD}C. Brans and R. H. Dicke, Phys. Rev. {\bf 124}, 925 (1961)
\bibitem{ban}N. Banerjee and D. Pavon, Phys. Rev. D {\bf 63}, 043504 (2001)
\bibitem{will}C.M. Will, Theory and Experiment in Gravitational Physics, (Cambridge University Press, 3rd edition, Cambridge, 1993)\\
C. M. Will, Liv. Rev. Rel. {\bf 9}, 3 (2005)
\bibitem{ban1}O. Bertolami and P. J. Martins, Phys. Rev. D {\bf 61}, 064007 (2000)\\
M. K. Mak and T. Harko, Europhys. Lett. {\bf 60}, 155 (2002)
\bibitem{ban2}W. Chakraborty and U. Debnath, arxiv:0807.1776v1
\bibitem{ban3}S. Das and N. Banerjee, Phys. Rev. D {\bf 78}, 043512 (2008)
\bibitem{fan}H. Farajollahi, M. Farhoudi, A. Salehi and H. Shojaie, Astrophys. Space Sci. {\bf 337}, 415 (2012)\\
H. Farajollahi and A. Salehi, JCAP {\bf 1011}, 006 (2010)\\
H.Farajollahi and A. Salehi, JCAP {\bf 07}, 036 (2011)\\
K. Saaidi, A. Mohammadi and H. Sheikhahmadi, Phys. Rev. D {\bf 83},
104019 (2011)
\bibitem{bis}Y. Bisabr, Phys. Rev. D {\bf 86} 127503 (2012)
\bibitem{far1} V. Faraoni, E. Gunzig and P. Nardone, Fund. Cosmic Phys. {\bf 20}, 121 (1999)
\bibitem{far}V. Faraoni, \emph{Cosmology in Scalar-Tensor Gravity} (Kluwer Academic Publishers 2004)
\bibitem{2}B. Schutz,  Phys. Rev. D {\bf 2} 2762 (1970)\\
J. D. Brown,  Class. Quant. Grav. {\bf 10}, 1579 (1993)
\bibitem{3}S. W. Hawking and G. F. R. Ellis, \emph{The Large Scale Structure of Space-Time} (Cambridge 1973, Cambridge University Press)
\bibitem{3a}O. Bertolami, C. G. Bohmer, T. Harko and F. S. N. Lobo, Phys. Rev. D {\bf 75}, 104016 (2007)
S. Nojiri, S. D. Odintsov and P. V. Tretyakov, Prog. Theor. Phys.
Suppl. {\bf 172}, 81 (2008)
\bibitem{4}
O. Bertolami and J. Paramos, J. Phys: Conf. Ser. 222. 012010\\
O. Bertolami, P. Frazao and J. Paramos, Phys. Rev. D {\bf 83},
044010 (2011)\\
O. Bertolami and A. Martins, Phys. Rev. D {\bf 85},
024012 (2012)
\bibitem{5}T. P. Sotiriou and V. Faraoni, Class. Quant. Grav. {\bf 25}, 205002 (2008)
\bibitem{6}V. Faraoni, Phys. Rev. D {\bf 80}, 124040 (2009)
\bibitem{k}J. Khoury and A. Weltman, Phys. Rev. D {\bf 69}, 044026 (2004)\\
J. Khoury and A. Weltman, Phys. Rev. Lett. {\bf 93}, 171104 (2004)
\bibitem{faulk} T. Faulkner, M. Tegmark, E. F. Bunn and Yi Mao,
Phys. Rev. D {\bf 76}, 063505 (2007)
\bibitem{wein}S. Weinberg, Gravitation and Cosmology, John Wiley
and Sons 1972
\end{thebibliography}
\end{document}